# Joint image reconstruction and segmentation of real-time cardiac MRI in free-breathing using a model based on disentangled representation learning

Tobias Wech[1,2], Oliver Schad[1], Simon Sauer[1], Jonas Kleineisel[1], Nils Petri[3], Peter Nordbeck[3], Thorsten A. Bley[1], Bettina Baeßler[1], Bernhard Petritsch[1], Julius F. Heidenreich[1]

[1]Department of Diagnostic and Interventional Radiology, University Hospital Würzburg, Würzburg, Germany
[2]Comprehensive Heart Failure Center Würzburg, Würzburg, Germany
[3]Department of Internal Medicine I, University Hospital Würzburg, Würzburg, Germany

**Correspondence**

Prof. Dr. Tobias Wech
Department of Diagnostic and Interventional Radiology
University Hospital Würzburg
Oberdürrbacherstr. 6, Würzburg, 97080, Germany
E-Mail: wech_t@ukw.de

**Funding Information**

This work was supported by the German Ministry for Education and Research under Research Grants 05M20WKA and 01KD2215A, and the Interdisciplinary Center for Clinical Research Würzburg under Research Grant F-437.

# Abbreviations

LV – Left Ventricle

RV – Right Ventricle

SV- Stroke Volume

ESV – End-Systolic Volume

EDV – End-Diastolic Volume

EF – Ejection Fraction

RR – interval between successive R peaks

MR – Magnetic Resonance

MRI – Magnetic Resonance Imaging

bSSFP – balanced Steady State Free Precession

ECG – Electrocardiography

ML – Machine Learning

NUFFT – Non-Uniform Fast Fourier Transform

DRL – Disentangled Representation Learning

SDNet – Spatial Decomposition Network

xSDNet – extended Spatial Decomposition Network

CG-SENSE – Conjugate Gradient Sensitivity Encoding

GPU – Graphics Processing Unit

FiLM – Feature-wise Linear Modulation

CS – Compressed Sensing

LRS – Low-Rank plus Sparse

CI – Confidence Interval

GSTF – Gradient System Transfer Function

VarNet – Variational Network

# Abstract

**Purpose**: To investigate image quality and accuracy of derived cardiac function parameters for a novel joint image reconstruction and segmentation approach based on disentangled representation learning, which is enabling cardiac cine imaging in real-time under free-breathing.

**Methods**: A multi-tasking neural network architecture incorporating disentangled representation learning was trained using simulated examinations based on data from a public repository and MR scans specifically acquired for developing this model. An exploratory feasibility study tested the obtained method in undersampled real-time acquisitions based on an in-house developed spiral bSSFP pulse sequence in eight healthy participants and five patients with intermittent atrial fibrillation. Images and predicted LV segmentations were compared to the reference standard of ECG-gated segmented Cartesian cine in repeated breath-holds and corresponding manual segmentation.

**Results**: On a 5-point Likert scale, image quality of the real-time breath-hold approach and Cartesian cine was comparable in healthy participants (RT-BH: 1.99 ± .98, Cartesian: 1.94 ± .86, p=.052), but slightly inferior in free-breathing (RT-FB: 2.40 ± .98, p<.001). In patients with arrhythmia, image quality from both real-time approaches was favourable (RT-BH: 2.10 ± 1.28, p<.001, RT-FB: 2.40 ± 1.13, p<.001, Cartesian: 2.68 ± 1.13). Intra-observer reliability was good (ICC=.77,95%-confidence interval [.75, .79], p<.001). In functional analysis, a positive bias was observed for ejection fractions derived from the proposed model compared to the clinical reference standard (RT-BH mean EF: 58.5 ± 5.6%, bias: +3.47%, 95%-confidence interval [-.86, 7.79%], RT-FB mean: 57.9 ± 10.6%, bias: +1.45%, [-3.02, 5.91%], Cartesian mean: 54.9 ± 6.7%).

**Conclusion**: The introduced real-time MR imaging technique is capable of acquiring high-quality cardiac cine data in 1-2 minutes without the need for ECG gating and breath-holds. It thus offers a promising alternative to the current clinical practice of segmented acquisition, with shorter scan times, higher patient comfort and increased robustness to arrhythmia and patient incompliance.

# 1 Introduction

Dynamic cine MR imaging is an essential building block for functional assessment in cardiac MRI, regardless of underlying pathology. In clinical routine, ECG-gated segmented sampling based on Cartesian readouts is currently the method-of-choice used at most centers, to obtain time-resolved views with sufficient spatial and temporal resolution as well as high signal-to-noise-ratio. A slight acceleration (R~2-3) based on parallel imaging is typically applied to shorten the patient's breath-holds, which are required for the assessment of approximately 15 spatiotemporal views (2D+t), adequately covering the entire heart from base to apex. A segmentation of the different compartments of the heart (myocardium, left ventricle (LV) and right ventricle (RV)) is subsequently performed for the end-systolic and the end-diastolic phase, to ultimately deduce functional parameters such as stroke volume (SV) or ejection fraction (EF). While the latter task was fulfilled in a manual fashion for a long time, semi-automatic or fully automatic tools [1] are increasingly being used for improved efficiency.

Segmented cine delivers depictions of the heart of high quality when patients are compliant, i.e. they can hold their breath for a few seconds, and heartbeats and ECG recognition are regular. If these requirements are not met - which is often the case - severe motion artefacts can corrupt the obtained images. The large temporal footprint of the method, which typically covers multiple cardiac cycles is responsible for this susceptibility. Therefore, efforts have been underway for years to establish alternative MR imaging techniques that provide data in real-time and free-breathing [2]. However, as these have not yet largely replaced classical segmented cine and are at best a fallback method for difficult cases, the necessary acceleration has so far been accompanied by a loss of quality, which radiologists are not willing to accept across the board.

Various strategies for reducing the acquisition time towards real-time cardiac MRI were implemented, covering different parts of the imaging chain. Read out trajectories, like echo planar imaging or spirals, which are *per se* more efficient than Cartesian sampling, were exploited for faster data sampling [3,4]. Image distortions due to off-resonance and gradient imperfections probably represent the biggest hurdles to be overcome along this way. In recent years, the latter problem has been countered very efficiently by correction methods based on a gradient response function [5–7]. Acquisitions, which are violating the Nyquist-criterion (a.k.a. "undersampling") became popular especially with the advent of parallel imaging [8] and were also exploited to waive ECG gating and breath-holds [9]. The introduction of compressed sensing to MRI [10] then enabled reconstructions of real-time measurements with a comparable temporal and spatial resolution as the segmented cine technique [11]. These so called "model-based" approaches were also combined with non-Cartesian trajectories, to utilize multiple acceleration potential at the same time [7,12,13]. Finally, data-driven methods, i.e. machine learning (ML) based reconstruction techniques, are increasingly being trained and applied as sophisticated models capable of transferring undersampled acquisition into images of even higher quality [14]. Many of these ML approaches can deliver real-time cardiac images at way lower latency than "hand-crafted" compressed sensing models, which frequently require many iterations of associated optimization algorithms. While this is desirable also for standard volumetry, it represents a mandatory prerequisite for advanced methods such as interventional cardiac MRI [15].

To work towards further improved efficiency and stability of MR-based cardiac volumetry, we propose a new technique for high-quality real-time dynamic cine imaging with fast image reconstruction and post processing. Data sampling is based on balanced steady state free precession (bSSFP) and undersampled spiral trajectories, corrected by a gradient system transfer function [7]. Acquisitions are reconstructed and segmented in a joint fashion using a novel multi-tasking neural network exploiting disentangled representation learning. The model is trained with simulated data derived from an open database as well as data acquired specifically for this project. The method was then evaluated in an exploratory feasibility study in healthy volunteers and patients with intermittent atrial fibrillation.

We published a conference abstract introducing a preliminary version of the approach based on radial data from an open database in [16]. For the cited work, segmented cine acquisitions in breath-hold were retrospectively undersampled to simulate acceleration. We now have trained and extensively validated the method with more efficient spiral (real-time) data, which were prospectively acquired for this purpose both under breath-hold and free-breathing.

## 2 Methods

### 2.1 Study population and ethics license

This prospective single-center study was approved by the local ethics committee under license ID 173/22_skpm. Written informed consent was obtained from each participant. Cardiac MRI was performed in 16 healthy study participants and 5 patients with cardiac arrhythmia. Eight of the healthy participants were used for training, the 8 remaining ones as well as the 5 patients were used to evaluate the method. Inclusion criteria was age > 18 years and for patients the condition of intermittent atrial fibrillation. Presence of arrhythmia was verified by ECG during the MRI. Exclusion criteria were any common contraindication to MRI.

### 2.2 MR pulse sequence and gradient waveform correction

A spiral balanced steady state free precession pulse sequence (bSSFP, see Fig. 1 & Table 1) was implemented on a 1.5 T whole body MR system (Siemens Magnetom Avanto Fit, Siemens Healthcare GmbH, Germany). A single real-time frame was acquired using 13 consecutive arms (TR=3.7ms), which were equally distributed in k-space and feature a temporal footprint of 48 ms. The undersampling factor with respect to Nyquist-sampling was approximately 5. An RF phase alternation of 180° ($\pm\alpha$) was fixed throughout all measurements and no frequency scout was acquired in addition to the shimming of the cardiac region of interest. Triangular rewinders were appended to the spiral read out to null the zeroth gradient moment.

To train the proposed artificial neural network, slightly altered acquisitions were additionally performed in healthy study participants: Data were sampled for the duration of a breath-hold of approximately 10s for each of the 10-15 short-axis slices. The trajectory patterns were rotated after repeating it for a period of time exceeding one RR interval. By acquiring 8 different orientations within one breath-hold, both the binning of real-time frames and the determination of fully sampled k-spaces - segmented across multiple RR intervals based on self-gating - was possible. This approach delivered matching pairs of undersampled and fully sampled data of the same cardiac phase, which were ultimately used to train the model described in 2.4 (please also refer to supplemental material for more details on the acquisition patterns). NUFFT (non-uniform fast Fourier transform) reconstructions of the former (undersampled real-time) represent the input of the proposed method (see "Input" in Fig. 2), NUFFT reconstructions of the latter (segmented cine), complemented by a manual segmentation thereof represent the two output images (see "Output 1" and "Output 2" in Fig. 2). All acquisitions with spiral gradient waveforms were corrected using a gradient system transfer function (GSTF, [17]) in post-correction. I.e., the GSTF was determined once for the system used, and was subsequently applied to obtain the k-space-trajectory actually played out by the scanner. This corrected trajectory was then used for all reconstruction throughout this study.

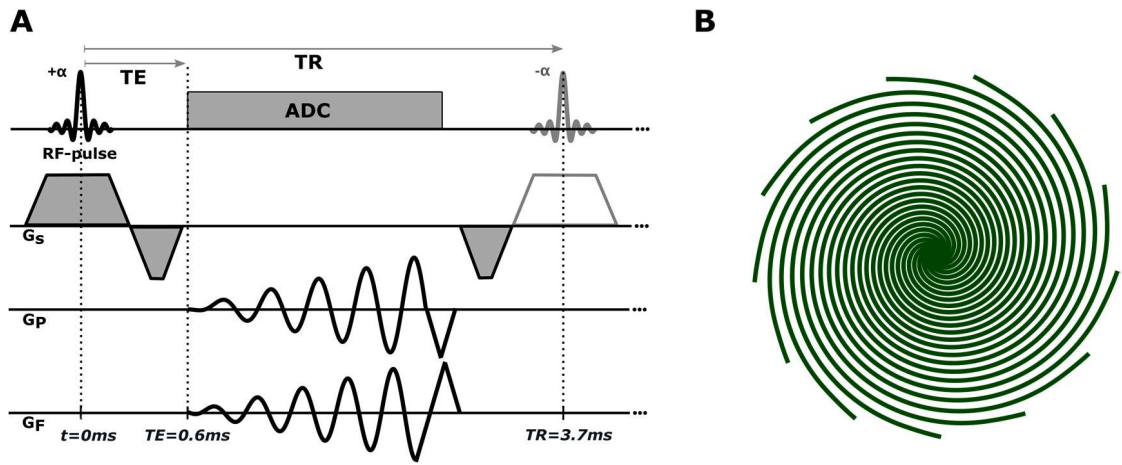

**Figure 1**: A) Spiral balanced steady state free precession MR pulse sequence. B) 13 consecutive spiral arms, equally distributed across $2\pi$, represent the k-space trajectory for one real-time frame (48 ms). To acquire training data for xSDNet in healthy volunteers, this pattern was repeated for a duration exceeding one RR-interval. Subsequently, the pattern was rotated to fill the largest gap in k-space. If data are acquired in breath-hold, and a sufficient number of orientations is sampled (>5 in this case), both real-time and segmented k-spaces (self-gating with DC signal) can be determined from the same acquisition.

## 2.3　Disentanglement

Disentangled representation learning (DRL, [18]) aims at factorizing data into several meaningful and disjoint (and thus independently varying) features. Compared to "black box" models, this more "human-like approach" to model data, potentially features increased interpretability, versatility, improved generalization and application robustness as well as multi-tasking capabilities. For medical imaging, Chartsias et al. [19] proposed a spatial decomposition network (SDNet) to disentangle radiologic depictions of the heart into an anatomical latent space and a modality latent space encoding contrast information. The approach was used to provide automatic image segmentation and style transfer of cardiac MRIs, as two exemplary applications of this intuitive and general model. In the following, we describe a network of similar nature, which is capable of simultaneously reconstructing and segmenting undersampled functional MR "real-time" investigations of the heart.

## 2.4 Model for joint reconstruction and segmentation

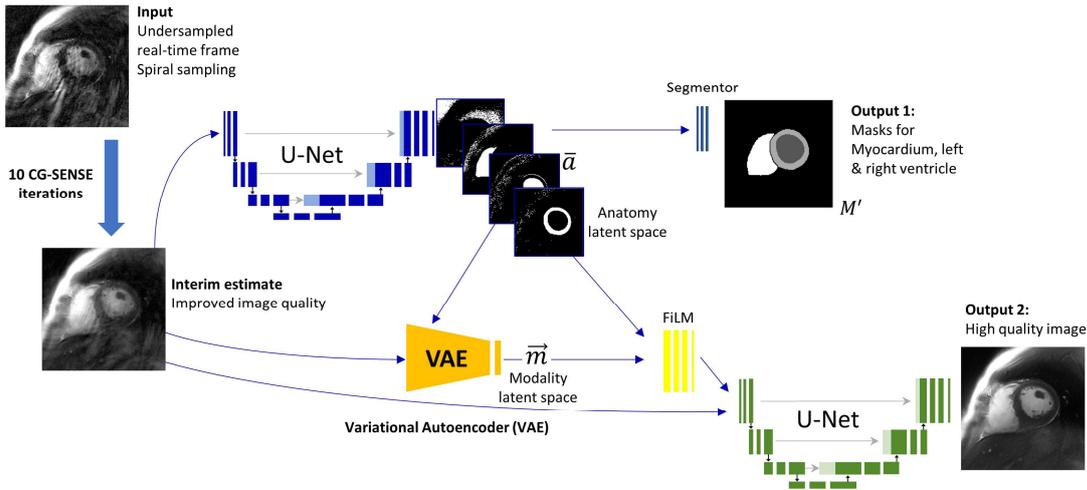

**Figure 2**: Illustration of the proposed xSDNet to provide joint reconstruction and segmentation of undersampled cardiac real-time frames.

The proposed joint reconstruction and post-processing strategy for cardiac real-time MRI is illustrated in Fig. 2. The input is represented by the undersampled (spiral) raw data acquired for one 2D real-time frame (48 ms footprint, see 2.2). These data are first subjected to ten iterations of a CG-SENSE [20] algorithm, incorporating GPU based convolution gridding [21] for fast run times. By exploiting the encoding capacity of the employed phased-array coil, this leads to an interim estimate of the real-time frame which already features less undersampling artefacts and therefore higher image quality than the naïve gridding reconstruction of the measured data. A temporal average image across multiple timeframes was used to determine coil sensitivities for this purpose [22].

The obtained interim 2D image is subsequently processed by a multi-compartment neural network which we refer to as e**x**tended **s**patial **d**ecomposition **net**work (xSDNet) in the following. The architecture is derived from the SDNet model proposed in [19] and was customized to both reconstruct and segment undersampled images with high quality. The upper path in Fig. 2 depicts the application of a U-Net [23] to determine an anatomical latent space, featuring eight anatomical factors represented by 2D Boolean matrices. This latent space builds the input to the so called segmentor, which is a shallow fully convolutional neural network consisting of three convolutional layers with batch normalization and LeakyReLu activation in between and a softmax classifier as final operator. The output is a mask for left and right ventricle as well as for myocardial tissue (Output 1). The lower path in Fig. 2 supplies the interim image to a variational autoencoder which is resulting in a "modality vector" of length eight, encoding the contrast information of the input. This modality latent space and the anatomy latent space from the upper path are subsequently fused by a decoding convolutional network featuring FiLM (Feature-wise Linear Modulation [24])-based normalization. This output was stacked with the interim estimate delivered by the CG-SENCE iterations to build the input of a final U-Net resulting in the high-quality reconstruction of the 2D real-time frame (Output 2). More detailed information on the architecture of the proposed model can be found in the supplementary material.

## 2.5   Training

xSDNet was trained using a combination of simulated (set 1) and self-acquired data (set 2). The former were determined using a subset of 96630 images from the "Kaggle Data Science Bowl Cardiac Challenge Data" [25], which were automatically segmented for LV, RV and myocardium using Bai's model [1]. Undersampled spiral raw data were additionally simulated for each image by superimposing artificial coil sensitivities (randomly varying for each sample) and applying a forward NUFFT operator [21], initialized by the spiral trajectory used for real-time imaging. Those data were subjected to ten CG-SENSE iterations to obtain the input of xSDNet (interim estimate, see Fig. 2). The two corresponding output labels are represented by the fully sampled image from the kaggle dataset (Output 2) and the segmentation masks (Output 1).

In addition, dedicated training data were acquired in eight healthy participants (set 2). 10-15 short axis slices were sampled in breath-hold as detailed in the second paragraph of section 2.2 and the supplementary material. Fully sampled images were obtained by applying the inverse NUFFT operator to 104 (= 8 × 13) self-gated spiral arms from eight consecutive heartbeats (Output 2). These images were segmented manually for LV, RV and myocardium by an MD candidate, who was trained for manually segmenting cardiac MR images by an expert (J.F.H) with seven years of experience in cardiac MRI. All segmentations used for this study were reviewed by the expert. Obtained segmentations represent the label for Output 1. Finally, interim estimates of the real-time frames - matched to the cardiac phases of the self-gated frames - were obtained by subjecting the 13 equidistant spiral arms (with a 48 ms footprint) to ten CG-SENSE iterations (input of xSDNet). 2010 input images with corresponding labels were obtained from the in-vivo study, which allow fitting the xSDNet model for the aimed target reconstruction and post processing task.

xSDNet was trained for 200 epochs. In each epoch, all data from set 2 were pooled with 282 randomly chosen samples from set 1, and, subsequently, a random horizontal and/or vertical flip combined with a random rotation by an angle $\in [0°, 90°, 180°, 270°]$ was performed for data augmentation. While remaining losses were used as proposed in the original SDNet [19], a perceptual loss [26] was used for the reconstruction path in our approach.

## 2.6   In-vivo study

The proposed method was tested in undersampled real-time spiral cine MRI at 1.5 T in eight healthy participants and five patients with intermittent atrial fibrillation. For better comparability with a clinical reference (see description below), data were first acquired in breath-hold. For patients, acquisition time was set to a breath-hold length of 4.5 s, resulting in 3-4 RR-intervals per slice, depending on heart rates. For healthy volunteers, a prolonged breath hold was used to preserve the possibility of determining a fully sampled (segmented) image series.

Subsequently, real-time data were acquired in free-breathing for the same slices, now in an end-to-end fashion. For each slice, data were sampled for 5.7 s. For a stack of 10-15 short axis views, the entire scan time consequently corresponds to 57-86 s for completely covering the LV. All data (breath-hold and free-breathing) were subjected to the joint reconstruction and segmentation by xSDNet and will be referred to as "test data" in the following.

Corresponding segmented Cartesian bSSFP cine series with ECG-gating were acquired in breath-hold as current clinical reference standard ("reference data"), again for the same slices as for the real-time investigations (see Table 1).

| Sequence parameter | Segmented Cartesian cine bSSFP | Undersampled spiral real-time bSSFP |
|---|---|---|
| Slice thickness [mm] | 8 | 8 |
| Flip angle [°] | 70 | 70 |
| Spatial resolution [mm] | 1.25x1.25 | 1.29x1.29 |
| FOV [mm] | 320x260 | 592x592 |
| Image Matrix [px] | 256x208 | 512x512 |
| TR [ms] | 3.94 | 3.70 |
| TE [ms] | 1.97 | 0.61 |
| Temporal resolution [ms] | 44±4 | 48 |
| Pixel Bandwidth [Hz/pixel] | 528 | 407 |
| GRAPPA | 2 | - |

**Table 1**: Acquisition parameters of the fully sampled segmented Cartesian bSSFP cine as well as the undersampled spiral real-time bSSFP pulse sequence. For the latter, the spatial resolution corresponds to $k_{max}$ (highest k-value sampled in k-space) of the spiral trajectory. The FOV represents the reconstructed FOV.

## 2.7 Comparison of disentanglement model with alternative methods

The test data reconstructed by xSDNet were compared to reconstructions of the same real-time data by CG-SENSE, $l1$-wavelet-based compressed sensing [10] (implementation of BART toolbox, [27]), low rank plus sparse (LRS, [28]) and a variational network [14]. It is important to note, that only LRS exploits redundancies in the temporal domain *t*, by applying a threshold to the series subsequent to a Fourier transform along *t* and an approximation of a corresponding spatiotemporal Casorati-matrix as low-rank. All other models reconstruct 2D frames individually and independently from other cardiac phases.

## 2.8 Expert reader study

Two expert readers (J.F.H., 7 years of experience in cardiac MRI, N.P. 3 years of experience) rated the real-time images as reconstructed by xSDNet and the corresponding ground truth images obtained by fully sampled segmented Cartesian acquisitions with ECG triggering. Experts were blinded to the reconstruction technique and the following items were rated: Artefacts in the blood pool, artefacts in the myocardium, sharpness of the endocardium, sharpness of the epicardium, depiction of dynamics, overall image quality. The items were rated on a 5-point Likert scale following the definitions for image quality assessment with 1: excellent, 2: very good, 3: good without impact on diagnostic quality, 4: fair with impact on diagnostic quality and 5: poor with strong impact on diagnostic quality.

## 2.9 Quantification of cardiac function

For comparison of cardiac function, endocardial and epicardial contours of the LV and contours of RV were segmented manually in Cartesian cine series (MEVISdraw v1.0, Fraunhofer MEVIS, Germany). Real-time data in breath-hold and free-breathing was segmented fully automatically by xSDNet. Since several cardiac cycles were covered in real-time, median values were determined for the automatically determined local maxima and minima. Ejection fraction was calculated from end-systolic and end-diastolic volumes.

## 2.10 Statistical analysis

Statistical analysis was performed with Python 3.12 using Pingouin [29] and Scipy libraries. For the expert reader study, ICC estimates and their 95% confident intervals were calculated based on a mean-rating (k=3), consistency and 2-way mixed-effects model. Friedman's test was used for comparison of xSDNet methods with Cartesian reference. Post-hoc pairwise comparison was performed with Wilcoxon ranked sum test. A p-value <0.05 was considered statistically significant.

# 3 Results

## 3.1 Comparison of disentanglement model with alternative methods

In Fig. 3, a systolic and a diastolic frame is depicted for the fully sampled ECG-gated Cartesian reference together with a "naïve" gridding reconstruction of undersampled spiral real-time acquisitions in breath-hold of a patient from the test set with atrial fibrillation. Real-time images were reconstructed by different methods as indicated (CG-SENSE, LRS, Variational Network, xSDNet). In the final image, the segmentation masks as provided by xSDNet are shown as a superposition on the image output.

The Cartesian reference (ECG-gated) shows clear motion artefacts, in both the systolic and the diastolic frame. This indicates that ECG-gating was not properly working for the irregular heartbeat, which corrupted the segmented acquisition and ultimately impaired the evaluation of cardiac dynamics. The naïve gridding reconstruction of the spiral real-time frames provides a good impression on how severe undersampling / aliasing artefacts would degrade image quality if no advanced reconstruction technique would be applied. In general, all images reconstructed from spiral data show increased blurring of fat in comparison to the Cartesian trajectory. CG-SENSE can significantly reduce the level of undersampling artefacts, however, residual streaks are apparent and SNR is sub-optimal. Using an additional $l1$-wavelet regularization further improved image quality, however, undersampling can still be recognized and typical wavelet-artefacts ("blocky-effect") can be seen in mild form. LRS features an overall sharper appearance, however, also slightly noisier as other techniques, except SENSE. The Variational Network has clearly higher SNR, while residual artefacts are slightly higher than for LRS. Overall best image quality was observed for the proposed xSDNet model. Both systolic and diastolic view represent sharp depictions of the respective cardiac phase, with high SNR and lowest artefact level. Moreover, the model simultaneously delivers an accompanying segmentation, which can be seen as an overlay in the final image of the figure. In supplementary material, a complimentary dynamic view can be seen, showing all frames of Fig. 3 (supplementary_video_1.gif). Motion artefacts within the ECG-gated Cartesian cine are even more clearly recognizable with respect to the static images, which is largely avoided by the real-time series. xSDNet shows high spatial and temporal sharpness and low artefact level.

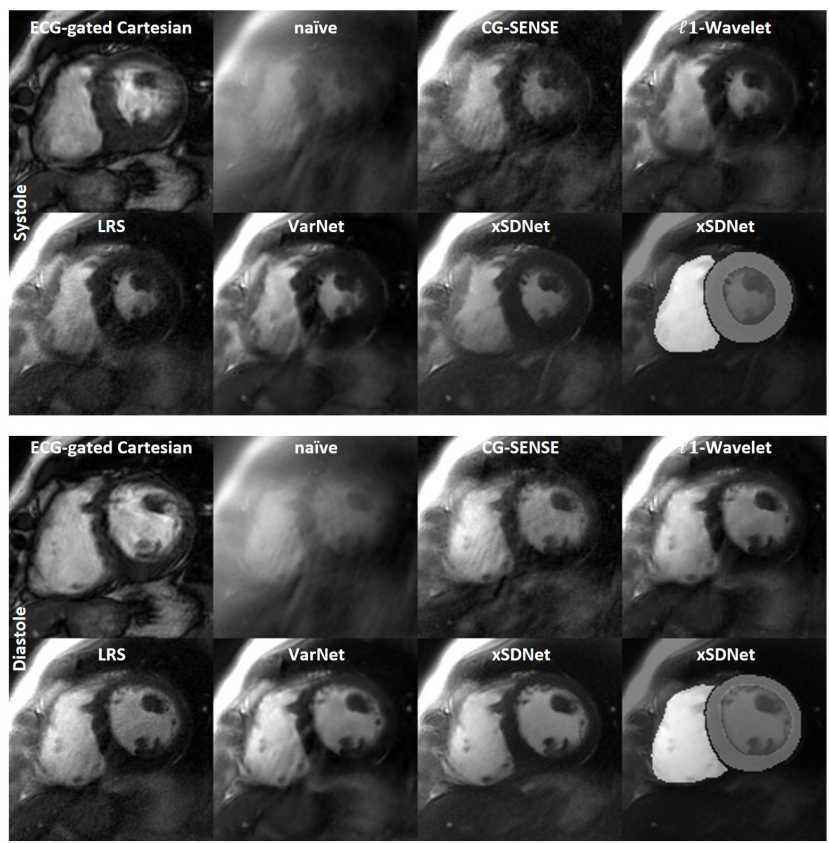

**Figure 3**: Performance of xSDNet in comparison to alternative reconstruction methods (breath-hold acquisitions) for a patient suffering from arrhythmia, both for a systolic and a diastolic cardiac phase. The image on top left, shows the result from the clinical method based on ECG-gated and segmented Cartesian bSSFP, respectively. All other images were reconstructed from undersampled spiral real-time acquisitions with a temporal footprint of 48ms. The final image shows the second output of xSDNet, i.e. a segmentation of the image, as overlay. See supplementary material for a dynamic view (supplementary_video_1.gif).

Figure 4 depicts results from real-time acquisitions in free-breathing in comparison to the clinical gold standard of Cartesian cine with segmented acquisition in breath-hold, both for a healthy volunteer with regular heartbeat and a patient suffering from arrhythmia. $l$1-wavelet based CS reconstruction of the real-time data are depicted as model-based reference. As for the acquisitions in breath-hold, high image quality was obtained by applying the xSDNet model. For the healthy participant, sharp depictions resulted from both reference cine and real-time approach. For the patient with irregular cardiac cycles, ECG-gated cine again suffered from distinct motion artefacts, while real-time frames reconstructed by xSDNet appear sharp, with sufficient SNR and low aliasing artefact level. CS reconstructions of the spiral real-time data equally provide a clear depiction of the dynamics, however, have higher overall residual undersampling artefacts compared to xSDNet. The findings are further confirmed by the dynamic views provided in the supplement for each participant presented in Fig. 4 (supplementary_video_2.gif and supplementary_video_3.gif). There, also for the healthy participant, some minor motion artefacts can be seen in the blood pool. For the patient, some instability of the segmentation performance can be seen as temporal jitter for myocardial tissue and more pronounced for the right ventricle.

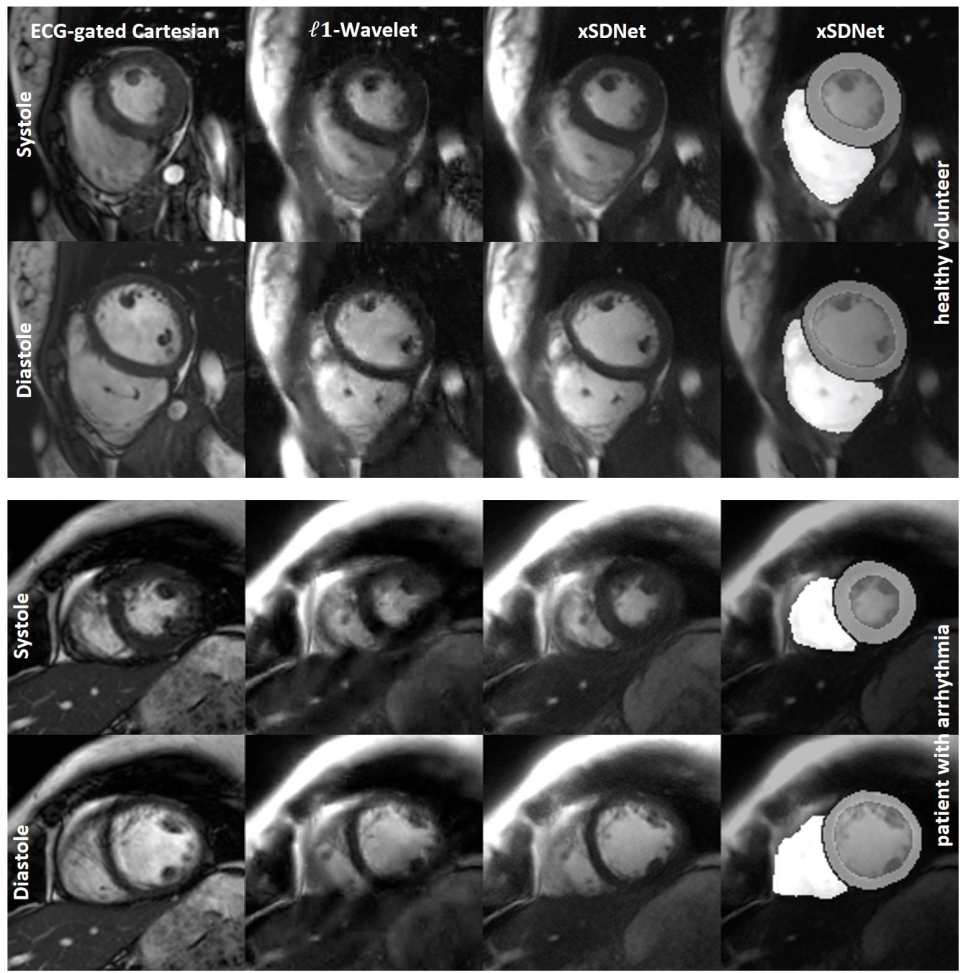

**Figure 4**: Performance of xSDNet based reconstruction and segmentation of real-time acquisitions in free-breathing in comparison to ECG-gated Cartesian cine in breath-hold and an l1-wavelet based compressed sensing reconstruction of the real-time data. See supplementary material for a dynamic view. Views differ slightly between xSDNet and cine due to breathing motion and significant delay between the two acquisitions. See supplementary material for dynamic views (supplementary_video_2.gif and supplementary_video_3.gif).

## 3.2 Expert reader study

In healthy participants, xSDNet in breath-hold and Cartesian cine achieved good ratings for overall image quality (xSDNet breath-hold: 1.99 ± .98; Cartesian: 1.94 ± .86; p=.052) with a slightly decreased impression in xSDNet free-breathing acquisition (xSDNet free-breathing: 2.40 ± .98, p<.001). In contrast, in patients with arrhythmia, both xSDNet acquisitions yielded better magnitude image quality compared to the reference (xSDNet breath-hold: 2.10 ± 1.28, p<.001, xSDNet free-breathing: 2.40 ± 1.13, p<.01, Cartesian: 2.68 ± 1.13). When focusing on the depiction of the dynamics of the heartbeat, realtime imaging with xSDNet has no pronounced benefit in healthy participants without arrhythmia (xSDNet breath-hold: 1.70 ± .91, free-breathing: 2.13 ± .99, Cartesian: 1.91 ± .84, Friedman's p=.09). In patients with arrhythmia however, xSDNet can maintain the high quality of dynamics (xSDNet breath-hold 1.87 ± 1.04, p<.001, xSDNet free-breathing: 2.22 ± 1.06, p<.001), yielding a relevant benefit compared to Cartesian imaging (2.69 ± 1.09, Figure 5). Additional image ratings are shown in supplementary Table 1. Intra-observer reliability was good (ICC = .77, 95% confidence interval [.75, .79], p<.001).

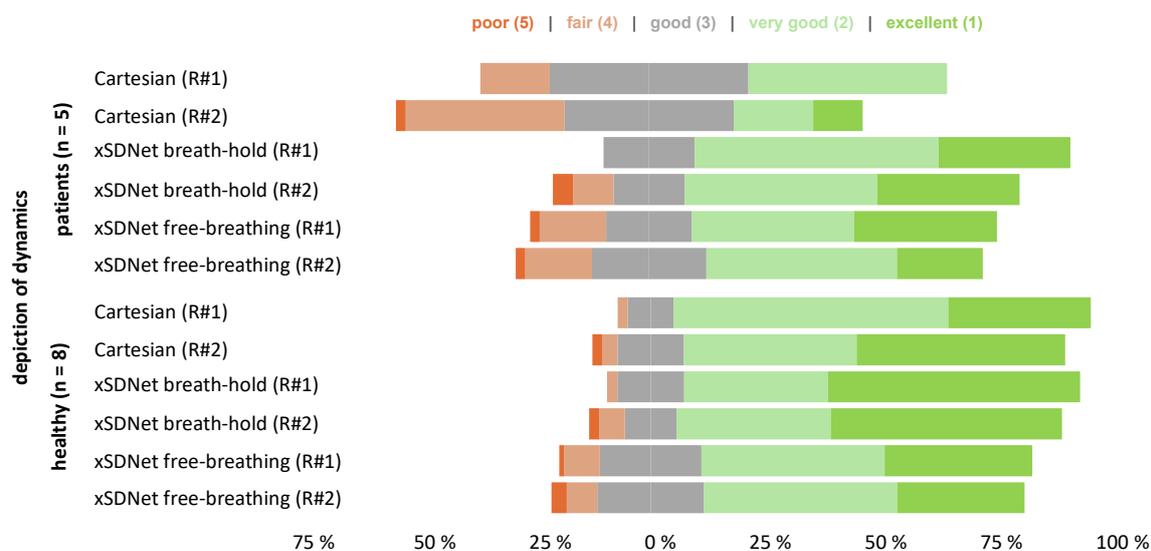

**Figure 5**: Results from the expert reader study. Distribution of ratings using a 5-point Likert scale are shown separately for both expert readers (R#1 & R#2) for the depiction of dynamics for xSDNet in breath-hold and free-breathing and Cartesian reference. While differences for the methods were marginal in healthy participants (n=8), a benefit for depiction of dynamics can be seen in patients with arrhythmia (n=5).

## 3.3 Cardiac function

One patient was excluded from functional analysis due to severe banding artefacts in apical slices throughout all acquisition methods which lead to poor manual and automatic segmentation. Overall, ejection fractions were widely comparable between methods for healthy participants and arrhythmic patients (Figure 6). Ejection fractions were slightly higher in breath-hold xSDNet real-time compared to Cartesian reference (bias +3.47%, limits of agreement [-.86, 7.79%], healthy: bias +4.02%, LoA [-.93, 8.96%], patients: bias +2.37%, LoA [-.15, 4.89%]. A smaller bias was observed in free-breathing (bias +1.45, LoA [-3.02, 5.91%], healthy: bias +.86%, LoA [-3.87, 5.59%]; patients: bias +3.01%, LoA [.22, 5.81%]). However, due to the exclusion of one additional subject due to bandings the data for the patient group in free-breathing is sparse (Figure 6, outlier is marked in red).

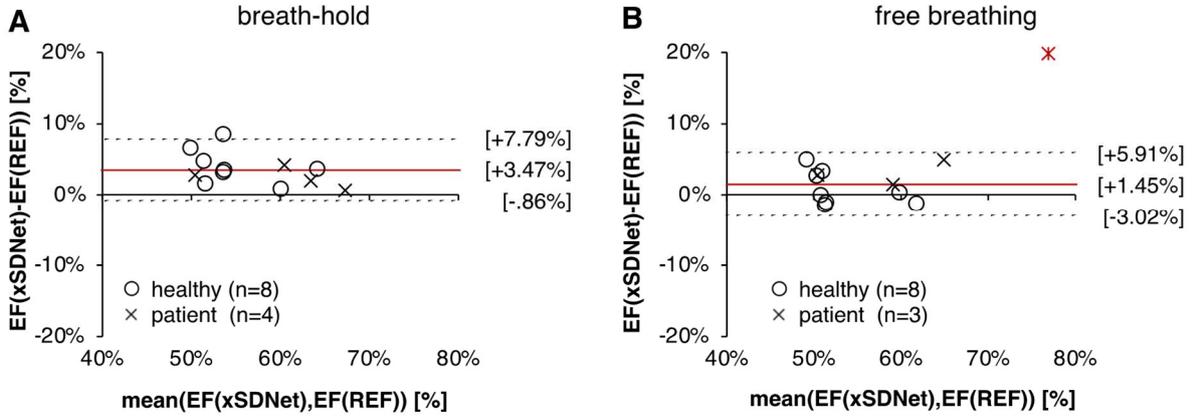

**Figure 6**: Bland-Altman plots for the ejection fractions determined by different methods. xSDNet represents the automatic segmentations of real-time frames as second output of the proposed neural network in breath-hold (A) and free-breathing (B). REF depicts the manual segmentations of the ECG-gated Cartesian reference. Bias is indicated in brackets and marked as red line. Upper and lower level of agreements are indicated and marked as dotted lines. In (B), one patient was additionally excluded from bias and agreement levels due to strong banding artefacts (outlier marked as red cross).

### 3.4 Speed of reconstruction

The prototype implementation of the reconstruction and segmentation pipeline processed images step by step (offline, and on separate workstations) and was not optimized for a clinical workflow. The iterations of CG-SENSE (# 10) for determining an interim estimate for one real-time frame took 2.10s on an NVIDIA Titan RTX GPU. Each 2D frame was then subjected to the xSDNet to determine high quality images and segmentation masks, which took 0.02s on an NVIDIA Titan X.

In addition, coil sensitivities were determined once for each real-time series of a specific slice. Gridding a temporal average on an NVIDIA Titan RTX took 0.33s, while the subsequent application of the algorithm by Walsh et al. [22], which adaptively determines coil maps was still CPU-based and took approximately 3.5s.

For a typical number of 80 real-time frames, the mean reconstruction time per frame can therefore be estimated at ~2.2s.

## 4 Discussion

We proposed a disentangled representation learning model xSDNet as a promising approach for the joint reconstruction and segmentation of undersampled spiral bSSFP motion studies of the heart. Coverage of the entire LV, from base to apex, can be acquired in 1-2 minutes under free-breathing, which represents a clearly shorter investigation with respect to the current state of the art of segmented acquisitions based on repeated breath-holds and typical investigation times of 10 min or more. In conjunction with the simultaneously delivered segmentation masks, implementation of xSDNet into clinical routine could significantly accelerate acquisition and post-processing of essential cardiac cine MRI.

*Image quality*

In our exploratory feasibility study, xSDNet reconstructions achieved very good to excellent image quality ratings. Image quality was equivalent in real-time breath-hold imaging and Cartesian cine, while ratings were slightly inferior in free-breathing imaging. Conversely, in patients with cardiac arrhythmia, both xSDNet acquisitions in breath-hold and free-breathing showed superior image quality in comparison with Cartesian segmented reference cine MRI (e.g. see Fig. 5). The latter exploits binning of data acquired across several heart beats, which results in excellent image quality for regular cardiac cycles and proper breath-holds. In patients who cannot hold breath or suffer from arrhythmia, this method, however, often delivers insufficient image quality and – maybe more severe – can corrupt derived functional parameters in a subtle manner, when the incorrect assignment of different cardiac phases leads to temporal blurring. In its current implementation, the temporal footprint of real-time frames is 48 ms and images are reconstructed individually, i.e. no temporal model is applied. This excludes any kind of temporal blurring for the images reconstructed by xSDNet, which is confirmed by the ratings of the item "depiction of dynamics". Solely breathing motion within the temporal window of 48 ms could lead to additional corruption of the data, which should, however, be largely negligible.

In general, the image quality of the real-time series reconstructed by xSDNet in terms of residual artefacts, sharpness, etc. was not inferior to the conventional fully sampled Cartesian cine acquisition. The initial application of ten conjugate gradient steps exploiting the physical information on coil sensitivities has a positive impact here and stabilizes the overall performance. Residual blurring due to off-resonance can be recognised in the accelerated images in the chest wall. If necessary, however, this can be further reduced using appropriate methods [5,30,31].

Nevertheless, it is conceivable to also exploit temporal redundancies within image reconstruction (and segmentation), with potential positive impact on the quality of real-time series (and masks). The advantages of processing frames individually, i.e. the possibility of processing frames in parallel and insensitivity to temporal overfitting, would, however, need to be waived with respect to the current implementation of xSDNet.

*Assessment of cardiac function*

In terms of image segmentation, results for the masks of the blood pool were sufficiently stable to detect local minima and maxima fully automatically, except for a small number of slices, where banding artefacts disturbed the semantic segmentation process of xSDNet. Derived ejection fractions were in good agreement with those from the clinical reference standard, with a slight positive bias for xSDNet at low variances of the deviations. For myocardial tissue and RV, the segmentations were somewhat more unstable, with a higher jitter in temporal observation and more frequent misclassification (see example shown in supplementary_video_3.gif). As the determination of myocardial mass or right ventricular ejection fraction from those would have at least required some kind of manual interaction, i.e. selection of frames, which were accurately segmented across all real-time frames, we have decided to dispense with this not fully automatic evaluation.

*Latency of image reconstruction and segmentation*
The current implementation allows a reconstruction time per frame of about 2.2s on average, which is clearly dominated by the CG-SENSE iterations applied as a pre-step. However, it should be noted that the algorithm was not optimized for overall run-time efficiency yet in this exploratory feasibility study. The following steps can be used to further reduce latency:
Even though estimating coil sensitivities has to be performed only once for a series, the run time of more than 3s of the current CPU implementation is still too long. We currently use images with full matrix size (512 × 512) as input for the algorithm proposed by Walsh et al. [22]. As coil sensitivities are smooth, however, this resolution is by far not necessary. Reducing the resolution by only a factor of four accelerates the total time for the estimation below 0.25s. As no temporal model is applied throughout the series, the CG-SENSE reconstruction of each frame can be run in parallel on a single or multiple GPUs. With a negligible run time of the neural network modules (0.02s per frame), this can reduce inference times well below 10s for a real-time series covering a few RR-intervals (2D+t), also without high-end hardware.

*Comparison with existing methods*
Recently, an alternative approach was proposed to reconstruct real-time cardiac MR data and automatically determine left ventricular EF [32]. In contrast to xSDNet, which exploits an integrated procedure to reconstruct and segment data in a multi-tasking fashion, the authors use a two-step approach, which first determines real-time images with a 3D U-Net based network and then perform the segmentation with another 2D U-Net.
In [15], real-time acquisitions based on spiral sampling were directly reconstructed using a modified fast convolutional neural network architecture, initially proposed for video denoising. For the focus of interventional studies, this allowed a latency of only 33ms per frame. To adjust xSDNet towards a comparable fast application, the initial CG-SENSE step could be waived. Here, however, we aimed at uncompromised image quality, which would have to be downgraded to some degree (e.g. lower spatial or temporal resolution, residual aliasing artefacts), if the physical information on coil sensitivities remains unconsidered and data are directly fed into the reconstruction and segmentation model.
In [33], a compressed sensing model was used to provide real-time cardiac imaging in free-breathing. Acquisitions across multiple heart beats were registered by a non-rigid correction of respiratory motion to allow for signal averaging and to improve slice positioning. In contrast to this approach, our intention was to not mix information from different RR-cycles. The median of multiple volumes was provided instead to allow a direct comparison of a synthetic RR-cycle from Cartesian cine MRI with our proposed real-time method.

*Limitations*
At 1.5T, the spiral bSSFP pulse sequence applied with a fixed phase between subsequent excitations of 180°, combined with a local shim as provided by the vendor of the MR system, resulted in hardly any banding artefacts inside the heart throughout all exams to acquire training data. For a small number of single slices of the test data, however, signal cancellations in myocardial tissue and blood pools were obtained, which especially had a major impact on the segmentation performance, both for manual and automatic processing. As xSDNet was applied offline so far, bandings could not be detected directly. An initial frequency scout would be reasonable to facilitate clinical transfer.

Even though a high number of simulated training sets (set 1) was added in each epoch, the number of training sets with 100% authentic contrast etc. (set 2) was comparably small, which was most likely the main reason for decreased segmentation performance, especially in apical slices fir myocardium and right ventricle. We are confident that increasing this number by ongoing studies on spiral cardiac MRI will further stabilize segmentation performance also for these compartments. In general, the initial findings of our exploratory feasibility study need to be further analysed in a subsequent more extensive study with higher statistical power.

The variational network used for comparison was trained with the spiral bSSFP sets acquired for training purposes only. As we rely on authentic raw data here, no external or simulated data was additionally used. However, the size of the training dataset is thus rather small and may have negatively influenced the performance of the reconstruction.

# 5 Conclusion

xSDNet applied to undersampled spiral bSSFP acquisitions in free-breathing enables real-time cardiac cine MRI with an image quality comparable to fully sampled Cartesian scans, an automatic determination of ejection fractions in agreement with the clinical reference standard, and significantly shorter scan times of 1-2 minutes for the entire heart. Higher patient comfort, reduced costs and increased robustness to arrhythmia and patient incompliance represent decisive advantages over current clinical practice.

## Acknowledgement

We thank Wenjia Bai et al. [1] for providing their segmentation model for scientific studies and Spyridon Thermos for providing a pytorch implementation of SDNet (https://github.com/spthermo/SDNet), which served as a baseline for our method.

## Supplemental files

supplementary_video_1.gif, supplementary_video_2.gif and supplementary_video_3.gif

Please contact us via <u>wech_t@ukw.de</u> to receive the video files.

## Software code

The source code of the presented xSDNet model will be made available upon publication in a peer-reviewed journal.

# Supplementary Material

Joint image reconstruction and segmentation of real-time cardiac MRI in free-breathing using a model based on disentangled representation learning

Tobias Wech, Oliver Schad, Simon Sauer, Jonas Kleineisel, Nils Petri, Peter Nordbeck, Thorsten A. Bley, Bettina Baeßler, Bernhard Petritsch, Julius F. Heidenreich

## 1. MR pulse sequence

The spiral balanced steady state free precession pulse sequence as illustrated in Fig. 1 of the main document is applied both in breath-hold and free breathing. The acquisitions in free breathing were solely used for real-time-depictions, which represents the intended and proposed application.

The acquisitions performed using a breath-hold were both used for further validation of real-time imaging <u>and</u> to obtain training data in healthy volunteers. For the latter, pairs of corresponding images, i.e. corresponding undersampled real-time and fully sampled "label" images, were required for the supervised training scheme applied. This was made possible, by a dedicated way of rotating the pattern shown in Fig. 1B of the main document along the temporal dimension (see Supp. Fig. 1):

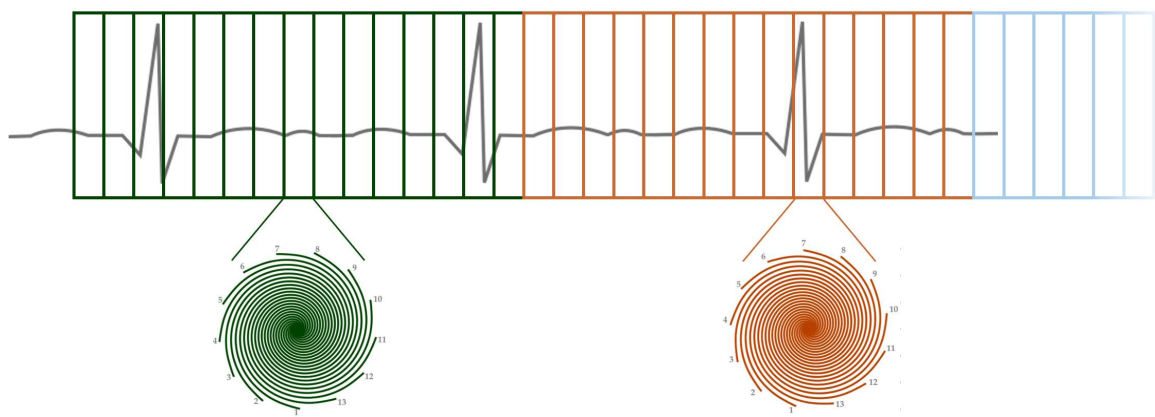

Supplementary Figure 1: Rotation of sampling pattern after a number of frames $n$.

The trajectory of one real-time frame consisting of 13 equally distributed spiral arms (green pattern in Supp. Fig. 1, corresponding to a temporal footprint of $48\ ms$) was played out repeatedly for a number of $n$ frames which are in sum exceeding the length of one RR interval (i.e. $n \cdot 48\ ms > RR$). Subsequently the pattern was rotated to fill the largest gaps (orange pattern) and was then again applied for the same number of $n$ frames. This was repeated 8 times, while each new pattern filled the largest gaps of the trajectory obtained by cumulative combination of all preceding patterns. Combining all 8 pattern ultimately delivers a trajectory which is "fully sampled".

By chosing $n \cdot 48\,ms > RR$, each cardiac phase (with a temporal accuracy of $\sim$ 48 ms) is sampled with each of the 8 patterns at least once, and as the acquisition was performed in breath-hold, the procedure then allows to determine fully sampled "segmented" frames and corresponding undersampled real-time frames for network training. $n$ thus had to be chosen individually for the current heart rate of the volunteers scanned for training.

When acquiring and reconstructing data in real-time only, there is no more need for building fully sampled frames in a segmented fashion for each heart-phase. In order to be able to determine temporal average images, however, we still applied the same pattern, but used considerably smaller values for $n$ (typically $n = 10$). As the start of the measurements is not triggered, the acquisition of $8 \cdot n$ (in our case ~80) needs to ensure the coverage of at least one RR cycle (in our case 3-4 cycles).

## 2. xSDNet architecture

The xSDNet architecture (see Fig. 2 in main document) is derived from the SDNet architecture introduced by Chartsias et al. [1]. The latter disentangles cardiac MRI images into spatial anatomical factors, which are capturing the physical structure of the heart etc., and non-spatial modality factors, which reflect how the image was acquired, and are thus independent from the actual anatomy of the individual patient (i.e. imaging modality, (MRI-) protocol, contrast agent, ...). The proposed original SDNet architecture was designed to analyze MRI mages in that respect, and thus to learn the two separate factors simultaneously. This comes with potential advantages of improved analysis of anatomical features, potentially leading to more accurate assessment of heart function and the ability to combine information from various imaging modalities (e.g., MRI with CT scan) by focusing on the common anatomical factors.

The authors highlight the general nature of the approach to factorize data into intuitive, meaningful and interpretable components, which inspired us to extend the concept to reconstruct and segment undersampled MR data, and thus to accelerate MRI. The ability of the method to derive binary anatomy maps (see Fig. 3a in [1]) was conserved in our case, even though undersampling / streaking artefacts are now superimposed on the input images. In particular, the maps could not only split up modality information, but also the artefacts originating from violating the Nyquist criterion (see $\bar{a}$ in the center of Figure 2, which represents maps of an inference of the model to patient data), and are thus comparable to those shown in [1]. The anatomy maps were used to derive segmentation maps by an additional shallow network.

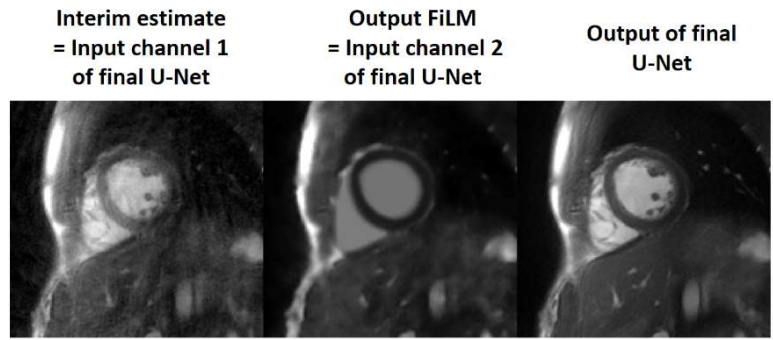

Supplementary Figure 2: The two input channels and the output channel of the final U-Net in Fig. 2 (illustrated in green) for an inference on test data.

Both the anatomy maps and the input image are used to derive the modality latent space by a variational autoencoder (identical architecture as in [1]), which is encoding information on contrast, thereby representing the "image modality". The two latent spaces (anatomy and modality) are then recombined by the "FiLM" module to obtain a reconstruction, as in the original SDNet. The obtained images – both in the original publication (see e.g. Fig. 1 and 3 in [1]) and also for our data (see Output FiLM in Supp. Fig. 2) – again represent both anatomical and modality features, however also appear somewhat "cartoonish" or oversimplified. However, for our application on accelerated scans, this estimated reconstruction also no longer features the undersampling artefacts, which are superimposed on the original image. We therefore used both the interim estimate and the output of the FiLM layer as inputs for the final U-Net, to deliver images free from artefacts but still containing detailed anatomical information (see image on the right in Supp. Fig. 2).

To study the benefit of this two-channel-approach, a comparison of the reconstruction performance of the xSDNet architecture with a radically ablated version, using the final U-Net only (trained with identical data, but just using the interim image as single input), was already presented in our ISMRM abstract [2]. There, a reader study confirmed, that xSDNet outperforms the benchmark U-Net for all acceleration factors applied (see Table 1 of the abstract and Fig. 3 for a visual impression).

## 3. Results: Detailed image ratings from the expert reader study

Additionally to overall quality of the magnitude images and the depiction of dynamics along the temporal domain, subjective image rating were performed to assess artefacts in bloodpool and myocardium as well as sharpness of the endocardial and epicardial contours.

| | all (n=13) | | | patients (n=5) | | | healthy participants (n=8) | | |
|---|---|---|---|---|---|---|---|---|---|
| | Cartesian | xSDNet BH | xSDNet FB | Cartesian | xSDNet BH | xSDNet FB | Cartesian | xSDNet BH | xSDNet FB |
| Artifacts in the bloodpool | 1.91±0.88 | 1.62±0,88 | 1.84±0.92 | 2.12±1.06 | 1.52±1.06 | 1.92±1.10 | 1.79±0.76 | 1.65±0.80 | 1.79±0.81 |
| Artifacts in the myocardium | 2.22±0.96 | 1.91±0.95 | 2.32±1.03 | 2.52±1.14 | 1.93±1.14 | 2.53±1.15 | 2.06±0.82 | 1.83±0.96 | 2.21±0.95 |
| Sharpness endocardium | 2.05±1.06 | 1.93±1.08 | 2.42±1.14 | 2.46±1.12 | 2.29±1.22 | 2.60±1.20 | 1.82±0.97 | 1.98±1.03 | 2.32±1.10 |
| Sharpness epicardium | 2.03±1.00 | 2.13±1.14 | 2.47±1.12 | 2.37±1.12 | 2.02±1.15 | 2.68±1.17 | 1.85±0.89 | 1.86±1.02 | 2.34±1.04 |
| Temporal dynamics | 2.12±1.02 | 1.76±0.98 | 2.39±1.07 | 2.69±1.09 | 1.87±1.04 | 2.22±1.06 | 1.91±0.84 | 1.70±0.91 | 2.13±0.99 |
| Magnitude image quality | 2.02±1.03 | 2.07±1.10 | 2.17±0.69 | 2.68±1.13 | 2.10±1.28 | 2.40±1.13 | 1.94±0.86 | 1.99±0.98 | 2.40±0.98 |

Supplementary Table 1: Data are shown as mean ± standard deviation for patients (n=5), healthy participants (n=8) and pooled (n=13). BH = breathhold, FB = free breathing.

## 4. Results: Real-time acquisitions in free breathing – comparison of image reconstruction based on xSDNet and Variational Network

An additional expert reader study compared the image quality of xSDNet reconstruction in free breathing with an alternative reconstruction approach with VARNET. As expected, xSDNet and VARNET reconstructions achieve similar ratings on the sharpness of endocardial/epicardial contours and the depiction of dynamics. With regards to artefacts in bloodpool and myocardium ratings were favourable for the xSDNet approach.

| | patients (n5) | | | |
|---|---|---|---|---|
| | Cartesian | xSDNet BH | xSDNet FB | VARNET FB |
| Artifacts in the bloodpool | 2.12±1.06 | 1.52 | 1.92±1.10 | 2.16±0.71 |
| Artifacts in the myocardium | 2.52±1.14 | 1.93 | 2.53±1.15 | 2.96±0.81 |
| Sharpness endocardium | 2.46±1.12 | 2.29 | 2.60±1.20 | 2.62±1.00 |
| Sharpness epicardium | 2.37±1.12 | 2.02 | 2.68±1.17 | 2.67±1.08 |
| Temporal dynamics | 2.69±1.09 | 1.87 | 2.22±1.06 | 2.23±0.94 |
| Magnitude image quality | 2.68±1.13 | 2.10 | 2.40±1.13 | 2.46±0.86 |

Supplementary Table 2: Data are shown as mean ± standard deviation for patients n=5 in free breathing.

## References for supplement